\pdfoutput=1
\documentclass[aps, reprint, twocolumn,longbibliography,nofootinbib,floatfix]{revtex4-1}
\usepackage[latin1]{inputenc}
\usepackage{graphicx}
\usepackage{amsmath,amssymb,amstext,graphicx,bm,booktabs,hyperref}
\usepackage{fancyhdr}
\usepackage{microtype}
\usepackage{bm}
\usepackage{xcolor}
\usepackage{multirow}
\usepackage{placeins}

\newcommand{\beginsupplement}{%
	\setcounter{table}{0}
	\renewcommand{\thetable}{S\arabic{table}}%
	\setcounter{figure}{0}
	\renewcommand{\thefigure}{S\arabic{figure}}%
	\renewcommand{\thesection}{*\Alph{section}}%
	\setcounter{equation}{0}
	\renewcommand{\theequation}{S\arabic{equation}}%
}

\begin{document}

\title{Quantum device fine-tuning using unsupervised embedding learning}

\author{N.M. van Esbroeck$^{1,2}$, D.T. Lennon$^1$, H. Moon$^1$, V. Nguyen$^1$, F. Vigneau$^1$, L.C. Camenzind$^3$, L. Yu$^3$, D.M. Zumb\"uhl$^3$, G.A.D. Briggs$^1$, D. Sejdinovic$^4$, and N. Ares$^1$}

\address{$^1$ Department of Materials, University of Oxford, Parks Road, Oxford OX1 3PH, United Kingdom}
\address{$^2$ Department of Applied Physics, Eindhoven University of Technology, PO Box 513, 5600 MB Eindhoven, The Netherlands}
\address{$^3$ Department of Physics, University of Basel, 4056 Basel, Switzerland}
\address{$^4$ Department Of Statistics, University of Oxford,
24-29 St Giles, Oxford OX1 3LB, United Kingdom}

\begin{abstract}

Quantum devices with a large number of gate electrodes allow for precise control of device parameters. This capability is hard to fully exploit due to the complex dependence of these parameters on applied gate voltages. We experimentally demonstrate an algorithm capable of fine-tuning several device parameters at once. The algorithm acquires a measurement and assigns it a score using a variational auto-encoder. Gate voltage settings are set to optimise this score in real-time in an unsupervised fashion. We report fine-tuning times of a double quantum dot device within approximately 40 min.


\end{abstract}

\date{\today{}}
\maketitle


\section{Introduction}

Electrostatically defined semiconductor quantum dots are intensively studied for solid-state quantum computation~\cite{Loss1998,Kloeffel2013,Hanson2007,Vandersypen2017}. Gate electrodes in these device architectures are designed to separately control electrochemical potentials and tunnel barriers~\cite{Camenzind2018,Camenzind2019}. However, these device parameters vary non-monotonically and not always predictably with applied gate voltages, making device tuning a complex and time consuming task. Fully automated device tuning will be essential for the scalability of semiconductor qubit circuits.


Tuning of electrostatically defined quantum dot devices can be divided into three stages. The first stage consists of setting gate voltages to create the confinement potential for electrons or holes. In our laboratory, full automation of this stage has been achieved as reported in Ref.~\cite{Moon2020}. The second stage, known as coarse tuning, focuses on identifying and navigating different operating regimes of a quantum dot device. Automated coarse tuning has been demonstrated using convolutional neural networks to identify the double quantum dot regime~\cite{Zwolak2019} and reach arbitrary charge states~\cite{Durrer2019}. Template matching was also used to navigate to the single-electron regime~\cite{Baart2016}. During this stage, virtual gate electrodes can be used to independently control the electrochemical potential of each quantum dot~\cite{Mills2019,Volk2019}. The third stage, referred to as fine-tuning, involves optimising a particular set of charge transitions. Previous work on automated fine-tuning focused on optimising the tunnel coupling between two quantum dots by systematically modifying gate voltages until this coupling converges to a target value~\cite{VanDiepen2018,Teske2019}. However, these approaches are restricted to a few device parameters and rely on calibration measurements.





Here, we demonstrate an automated approach for simultaneous fine-tuning of multiple device parameters, such as tunnel rates and inter-dot tunnel coupling. Our approach is based on a variational auto-encoder (VAE). In particular, we focus on double quantum dot devices. Electron transport through these devices is typically presented as a charge stability diagram, displaying the current flowing through the device as a function of two gate voltages. Bias triangles are regions in the stability diagram, for which current flow is allowed through a double quantum dot device under a bias voltage~\cite{Vanderwiel2002}, and reveal most of the device parameters. Our algorithm aims at optimising various bias triangle characteristics commonly associated with favourable device parameters, as done by humans when tuning these devices. The VAE compresses training data displaying bias triangles to a lower-dimensional space, called the latent or embedding space. In this latent space, a human expert identifies target locations corresponding to bias triangles in the training set which exhibit favourable transport characteristics. The algorithm acquires a measurement displaying bias triangles and assigns to these bias triangles a location in latent space. The distance between this location and the chosen target locations is used by the algorithm as a basis to score the measurement, and this score is used to optimise the gate voltage settings in real time.




We have previously shown that VAEs signifcantly improve the efficiency of quantum dot measurements~\cite{Lennon2019}. We have now, for the first time, used a VAE to fine-tune a double quantum dot device by locally optimising transport features in a completely automated manner. Without requiring any prior knowledge of the device architecture, we are able to fine-tune several device parameters at once. 
\section{Device and overview of the algorithm}

\begin{figure}[ht]
\centering
  \includegraphics[width=1\linewidth]{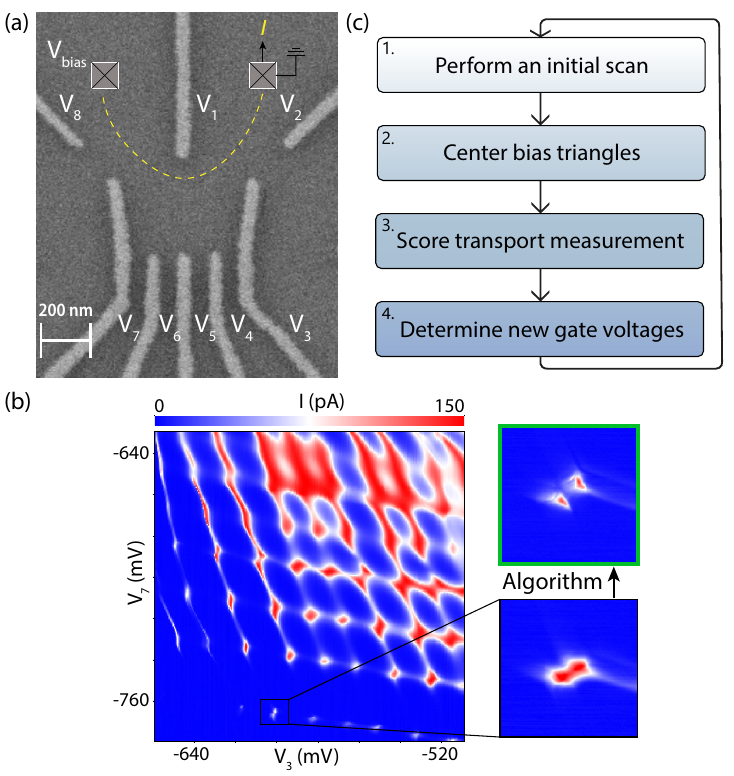}
  \caption{Overview of the quantum dot device and algorithm. (a) Scanning electron microscopy image of a device lithographically identical to the one measured. A bias voltage $V_{bias}$ is applied between two ohmic contacts to drive a current $I$ through the device. Gate voltages $V_1$ to $V_{8}$ define and control the double quantum dot. (b) Current as a function of gate voltages $V_3$ and $V_7$, with $V_{bias}=0.2$ mV. In this stability diagram, at the cross-points of the hexagonal lattice representative of the double quantum dot regime, bias triangles are observed. The zoom in shows a pair of bias triangles that requires optimisation (bottom). This pair of bias triangles is displayed after optimisation (top), showing how the triangular shapes can now be distinguished. (c) Schematic overview of the fine-tuning algorithm. In each iteration an initial low resolution stability diagram displaying bias triangles is acquired (1). Subsequently, the bias triangles are centered in a gate voltage window using blob detection (2). In this window, the algorithm performs a high resolution measurement which is scored by the VAE (3). Based on the VAE score, the decision model proposes a new gate voltage configuration (4).} 
  \label{fig1}
\end{figure}

We demonstrated our fine-tuning algorithm on a lithographically defined double quantum dot device. The device comprises a GaAs/AlGaAs heterostructure confining a two-dimensional electron gas (2DEG). Quantum dots are defined by Ti/Au gate electrodes which are patterned on top of the heterostructure (Fig.~\ref{fig1}a). DC voltages $V_1$ to $V_{8}$ are applied to these gate electrodes. A bias voltage $V_{bias}$ determines the flow of current $I$ through the device. A stability diagram for our double quantum dot device is shown in Fig.~\ref{fig1}b. All measurements were performed at approximately $20$~mK. The stability diagram exhibits bias triangles, which reveal device parameters such as charging energies, tunnel coupling to the left and right electrodes and inter-dot tunnel coupling. The shape, sharpness and brightness of bias triangles are related to those device parameters and are thus used to guide device tuning. 

Our algorithm follows a similar approach to device tuning by humans. It consists of four major steps (Fig.~\ref{fig1}c). In each iteration, an initial low resolution stability diagram is acquired to center a pair of bias triangles. Next, a high resolution measurement of the bias triangles is performed and scored by the VAE. Based on this score, the set of gate voltages is determined for the next iteration.

\section{VAE implementation}
\label{sec:vae_implementation}



The VAE consists of an encoder and decoder, both embodied in neural networks~\cite{Kingma2014}. The encoder $q_{\phi}\left(\bm{z}\vert \bm{x} \right)$ maps input data $\bm{x}$ to a low-dimensional latent vector $\bm{z}$ which is real-valued. The decoder $p_{\theta}\left(\bm{x}\vert \bm{z} \right)$ maps a latent vector to a reconstruction $\hat{\bm{x}}$. The parameters of the encoder and the decoder neural networks are $\phi$ and $\theta$, respectively. The VAE is a generative model; it seeks to preserve the maximum amount of information during the encoding process so that input data can be reconstructed with minimal error during the decoding process. During a training phase, $\phi$ and $\theta$ are iteratively updated to minimize a loss function. The loss function is given by a reconstruction error $\mathcal{L}_{rec}$, which penalises the networks for producing reconstructions that are dissimilar from the input data, and a regularisation term $\mathcal{L}_{reg}$, which enforces input data with similar characteristics to be encoded in close proximity in latent space. The reconstruction error and the regularisation term have weights $\alpha$ and $\beta$, respectively.

We implement Factor-VAE~\cite{Kim2018}, an adaption of VAE that seeks to generate a latent space in which each dimension corresponds to a unique characteristic of the input data. The Factor-VAE framework assumes that there are underlying independent factors associated with the data. If fully disentangled, each of those factors can be identified with a dimension in latent space. By using a Factor-VAE, we aim to generate a latent space in which each dimension is associated with a single bias triangle characteristic, such as size or brightness. In this way, the distance in latent space to a target location results in a good metric to score acquired measurements. 


The loss function of Factor-VAE includes a total correlation term which encourages the distribution of embeddings $q_{\phi}\left(\bm{z}\right)$ to be disentangled. It is given by: 

\begin{multline}
 \mathcal{L}_{Factor-VAE}= \mathcal{L}_{rec}+\mathcal{L}_{reg}\\
 -\gamma D_{KL}\left(q_{\phi}\left(\bm{z}\right) \vert\vert \textstyle{\prod_{j}} q_{\phi} \left(z_j\right) \right)
\label{eq:LFVae}
\end{multline}
where the total correlation term is given by $D_{KL}\left(q_{\phi}\left(\bm{z}\right) \vert\vert \textstyle{\prod_{j}} q_{\phi} \left(z_j\right) \right)$, i.e. the Kullback-Leibler divergence between the distribution of embeddings $q_{\phi}\left(\bm{z}\right)$ and the product of the distribution of embedding components $\textstyle{\prod_{j}} q_{\phi} \left(z_j\right)$, with the index $j$ corresponding to the $j$th latent space dimension. The total correlation loss term has a weight $\gamma$. Since this term is intractable, it is estimated using a discriminator $D\left(\bm{z}\right)$. The discriminator is trained to classify between non-factorial and factorial samples, i.e.~that its input is a sample from $q_{\phi}\left(\bm{z}\right)$ rather than from $\textstyle{\prod_{j}} q_{\phi} \left(z_j\right)$.

The training set for the Factor-VAE was collected from a device which differs considerably in material, architecture and transport regime from the device used to demonstrate the performance of the algorithm, evidencing its generality. The VAE was trained using 2253 sets of bias triangles, measured on a double quantum dot defined in a Ge/Si core-shell nanowire~\cite{Camenzind2019_thesis,Froning2018}. In order to increase the robustness of the VAE, simple data augmentation techniques were applied. Data augmentation included translation, rotation, mirroring, Gaussian noise and random contrast, resulting in a total training set of 8732 stability diagrams of pixel resolution 32 x 32. The dimension of the latent space was set to 10, as in Ref.~\cite{Kim2018}, given the similar structure of input data. We tried multiple combinations of weights $\alpha$, $\beta$ and $\gamma$ to achieve the optimal VAE performance, which was found emperically for $\alpha=34$, $\beta=1$ and $\gamma=1$. 

\section{Score metric}

The score metric used by the algorithm is given by the distance between the latent space representation $\bm{z}$ of an input stability diagram and the latent space representation $\{{\tilde{\bm{z}}}\}$ of a set of target inputs. Note that the loss function is used during training, while the score metric is used for optimisation. A measurement acquired by the algorithm is assigned a low (high) VAE score if its representation in latent space is near to (far from) the targets in latent space. Embeddings that are close together in latent space have similar $\bm{z}$, implying that the original inputs can be generated using similar underlying variables. As a result, bias triangles that are assigned a low score possess similar characteristics to the target bias triangles. 

The target bias triangles are chosen from the unaugmented training set by a human expert who recognises in these triangles the characteristics indicative of favourable quantum dot parameters. The targets are augmented using the same augmentation techniques as described in~\ref{sec:vae_implementation}. Augmentation of 30 selected targets resulted in a total target set of size 360. 

\begin{figure}[ht]
\centering
  \includegraphics[width=1\linewidth]{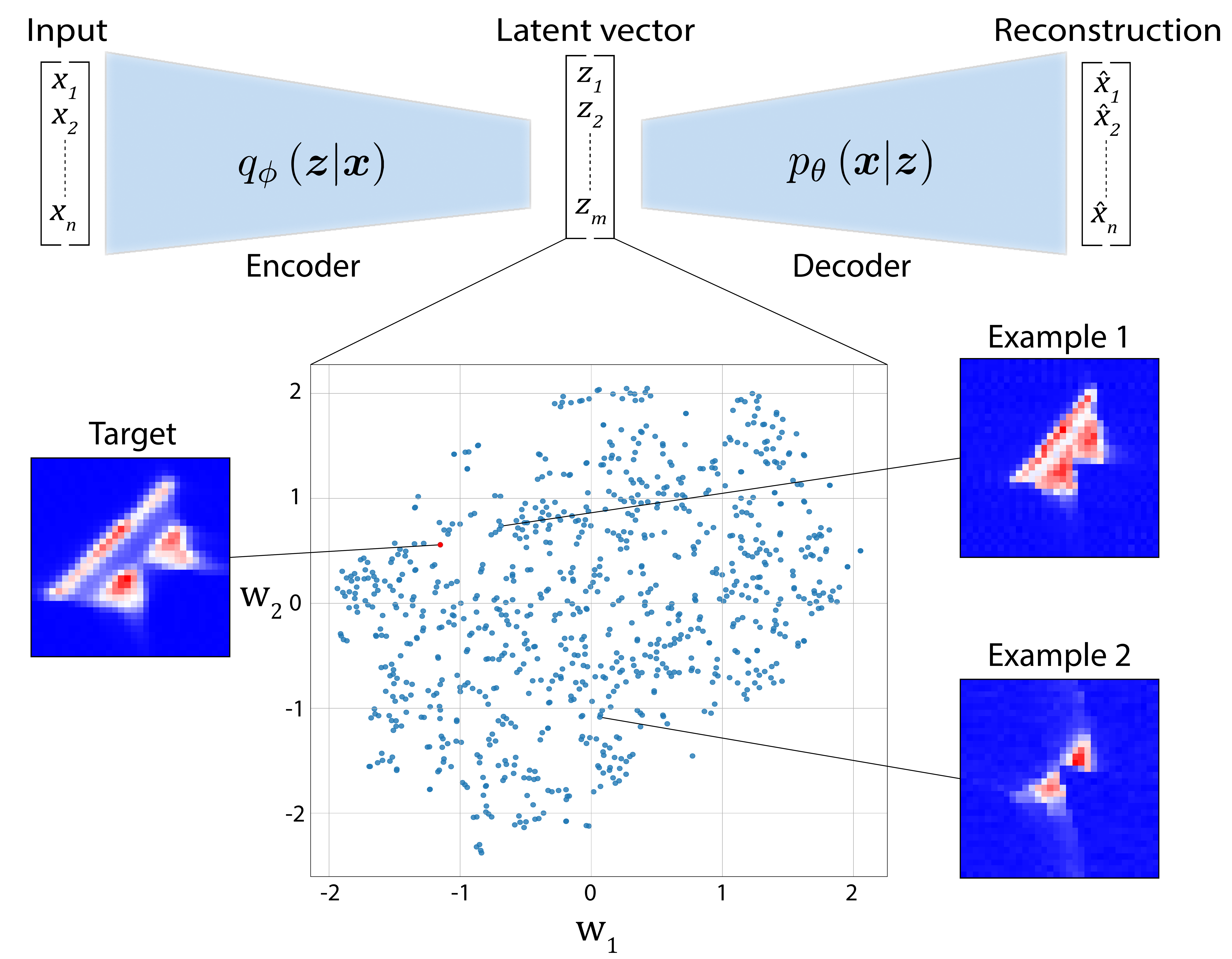}
  \caption{Schematic overview of the VAE. The VAE consists of an encoder and decoder. The encoder $q_{\phi}\left(\bm{z}\vert \bm{x} \right)$ compresses input stability diagrams to a lower-dimensional latent space. The decoder is denoted by $p_{\theta}\left(\bm{x}\vert \bm{z} \right)$ and maps vectors in latent space $\bm{z}$ to the distribution of input data. In this way, the input vector $\bm{x}$, for which each element is the brightness of one pixel in the stability diagram, is transformed into a reconstruction vector $\hat{\bm{x}}$. In order to visualise the ten-dimensional latent space, t-SNE is applied for dimensionality reduction~\cite{vanDerMaaten2008}. The resultant two-dimensional latent space is described by a vector $\bm{w}$. Each dot represents the embedding of an input stability diagram. The embedding location of one of the target inputs is highlighted in red. It is expected that embeddings which are close to each other in latent space are generated by input data with similar characteristics. Test example 1, with similar characteristics to the target, can be found in close proximity to the target, whereas test example 2 is further away in latent space.
  }
  \label{fig2}
\end{figure}

In Fig.~\ref{fig2} the latent space of the trained VAE is shown, and the embedding locations of example target and training inputs are indicated. A full plot of the latent space of the trained VAE with original input stability diagrams is shown in the Supplementary Material.

To write the expression for the score $S_i$, where $i$ denotes the $i$th input measurement, we use the latent vector $\bm{z}^i$ produced by the encoder for this measurement. The output of the encoder is assumed to follow a multivariate Gaussian with diagonal covariance structure: $q_{\phi}\left(\bm{z}\vert \bm{x} \right)= \mathcal{N}\left(\bm{z}; \bm{\mu}, \mathrm{diag}\left( \bm{\sigma}^2\right)\right)$, where the mean $\bm{\mu}$ and variance $\bm{\sigma}^2$ are outputs of the encoding network. Considering two independent normal distributions in latent space, the expectation value of the squared distance between the distributions is given by: 

\begin{multline}
   d\left(\bm{z}^1,\bm{z}^2\right)=\mathbb{E}\left(\|{\bm{z}^1-\bm{z}^2}\|^2\right)= \|\bm{\mu}_{\bm{z}^1}-\bm{\mu}_{\bm{z}^2}\|^2 \\+ \bm{\sigma}^2_{\bm{z}^1}+
 \bm{\sigma}^2_{\bm{z}^2}
 \label{eq:distance}
\end{multline}

For each input measurement embedded in latent space $\bm{z}^i$, (\ref{eq:distance}) is used to determine the distance in latent space to target input $\tilde{\bm{z}}^{j}$. The final score $S_i$ consists of the average of the distance to its $k$ nearest targets: 

\begin{equation}
 S_i=\frac{1}{k}\sum^k_{\tilde{\bm{z}}^{j} \in A_k} d\left(\tilde{\bm{z}}^{j}, \bm{z}^i \right)
\label{eq:score}
\end{equation}
where $A_k$ is the set of $k$ targets closest to $\bm{z}^i$ in latent space. In this way, optimal tuning corresponds to a low score. We found that for $k=3$, the score metric produced a robust ranking of the training inputs in terms of their similarity to the targets.

\section{Optimisation}

The optimisation starts from the device tuned to the double quantum dot regime so that at least one pair of bias triangles is identified, for which we used the algorithm presented in~\cite{Moon2020}. After acquiring the initial low resolution stability diagram, the bias triangles are centered using Laplacian of Gaussian (LoG) blob detection. In computer vision, blob detection techniques aim to detect bright regions on dark backgrounds or vice versa~\cite{Lindeberg1993}. In Fig.~\ref{fig1}b, the two examples of bias triangles displayed were centered with this novel approach. Once the triangles are centered, the high resolution scan (32 x 32 pixels, 17 x 17 mV) is acquired and evaluated using the VAE distance score metric. Based on the outcome value of $S_i$, a decision model sets the gate voltage configuration for the next stability diagram measurement. This process is iterated until the bias triangles are optimised in terms of characteristics such as shape, sharpness and brightness.  




\begin{figure}[ht]
\centering
  \includegraphics[width=0.8\linewidth]{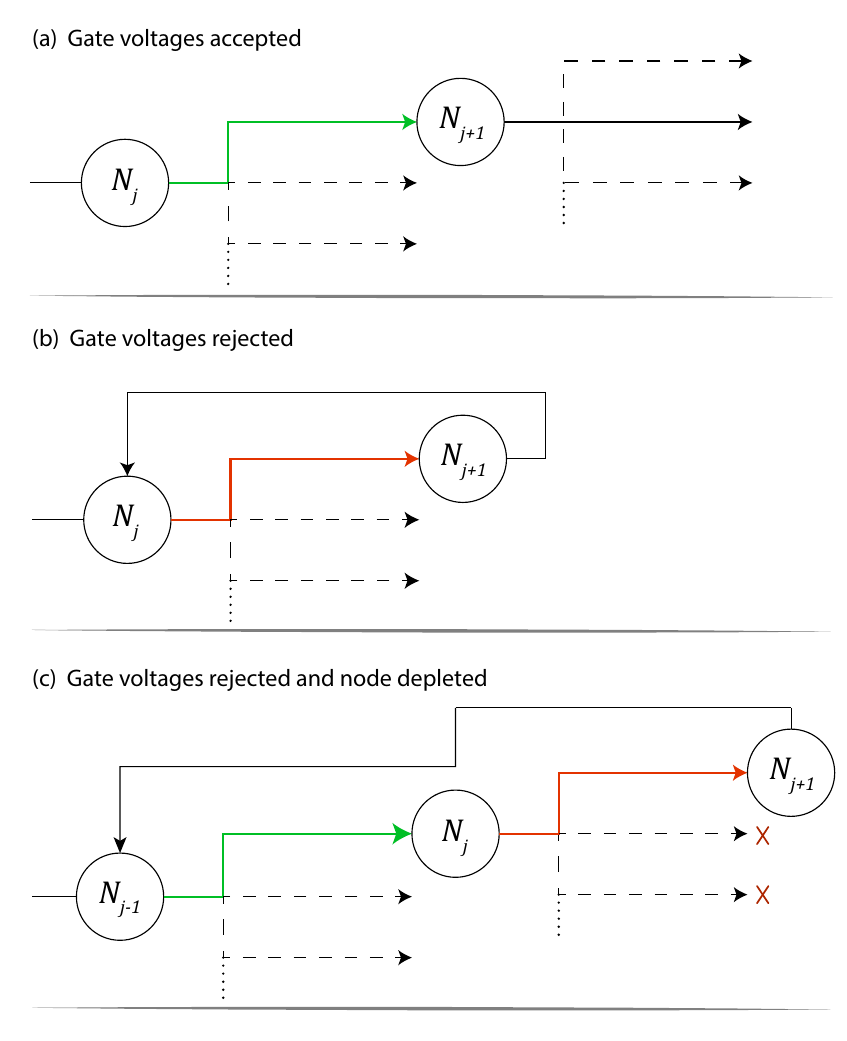}
  \caption{Overview of the decision model. The set of applied gate voltages number $j$ is indicated by $N_j$. The branched arrows represent the different gate voltage adjustment options, which are changes of $\pm \Delta V$ in every gate electrode to be tuned. In this figure, $N_j$ represents the best scored gate voltage configuration obtained after a number of iterations. (a) Score $S_i$ corresponding to a new configuration $N_{j+1}$ is lower than at $N_{j}$, so the gate voltage change is accepted. For $N_{j+1}$, a new random gate voltage branch is selected and explored. (b) Score $S_i$ corresponding to a new configuration $N_{j+1}$ is higher than at $N_{j}$, so the gate voltage change is rejected. For $N_{j}$, one of the remaining gate voltage branches is randomly selected and explored. (c) If all possible gate voltage configurations are rejected the algorithm returns to the closest previously accepted gate voltage node that has unexplored branches. At this configuration, a gate voltage branch is randomly selected and explored.}
  \label{fig3}
\end{figure}

The decision model for proposing gate voltage configurations is illustrated in Fig.~\ref{fig3}. Node $N_j$ represents the set of gate voltages ${\{V^j_1,V^j_2,...V^j_8\}}$ applied by the algorithm. In each iteration, one gate electrode is selected at random, and the voltage applied to this electrode is modified by a fixed amount $\pm \Delta V$. Therefore, the algorithm chooses between a number of branches equal to twice the number of gate electrodes to be tuned. We chose $\Delta V=2$ mV based on human experience in tuning similar devices. After centering and acquiring a high resolution measurement of the resulting bias triangles, the value of $S_i$ determines the algorithm's decision. If $S_i$ is lower than the previously best (lowest) scored bias triangles, the gate voltage change is accepted, leading to a new gate voltage configuration $N_{j+1}$. Conversely, if $S_i$ is higher, the gate voltage change is rejected and the gate voltage setting returns to its previous configuration. In this case, the rejected gate voltage change will be an excluded branch in the random selection corresponding to the next iteration. Branches that lead to the reversal of the latest accepted gate voltage change are excluded too. It is possible that all gate voltage branches become depleted, in which case the decision model returns to the previously accepted gate voltage configuration with unexplored branches.




\section{Experimental demonstration}

\begin{figure*}[ht]
\centering
\includegraphics[width=\linewidth]{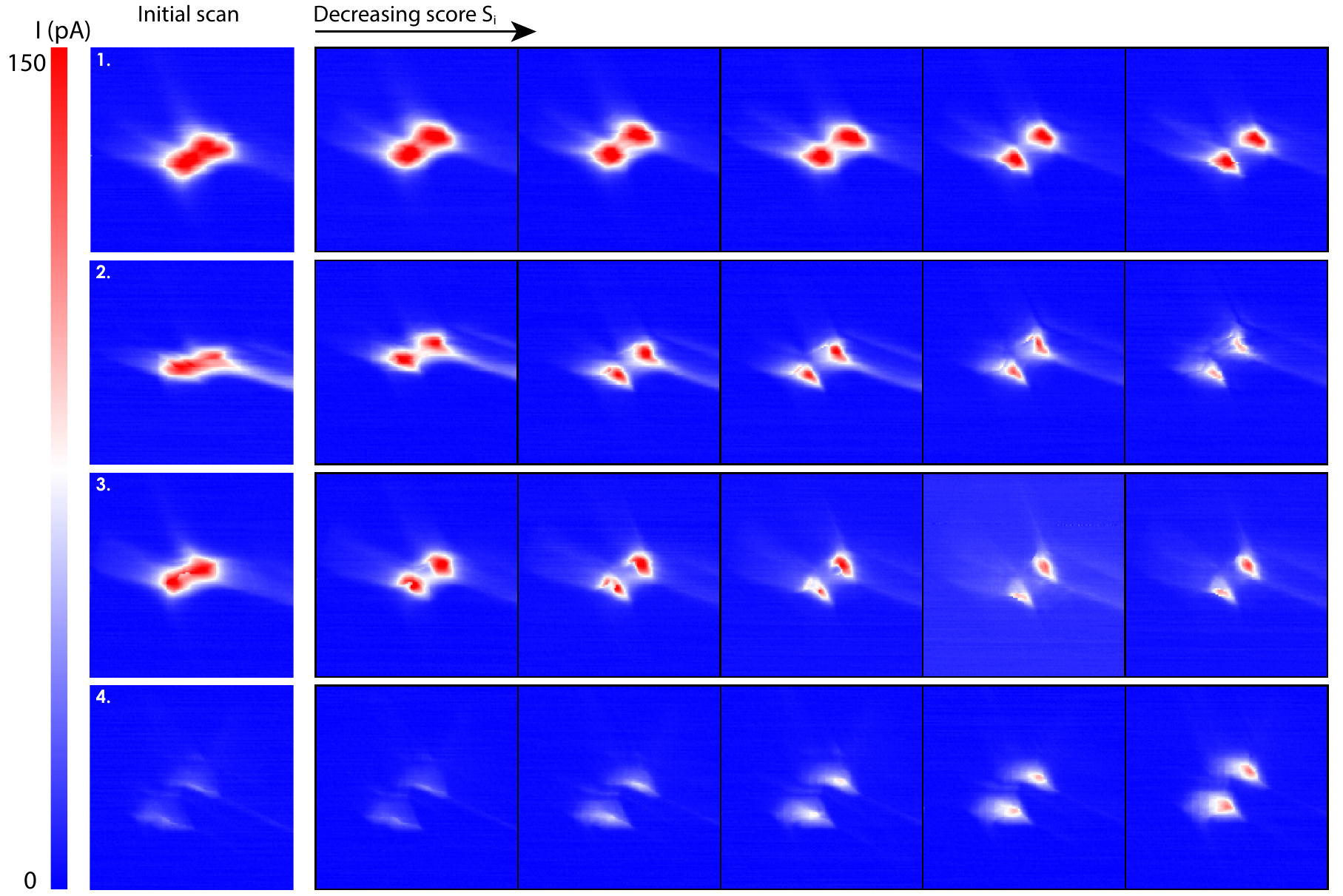}
  \caption{Experimental demonstration of the algorithm. The first column shows four different pairs of bias triangles before the algorithm was run. Each row displays each of these bias triangles at selected iterations of the algorithm. For all these iterations, the applied gate voltage change led to a decrease in score $S_i$. All measurements were performed with $V_{bias} = 0.2$ mV. The stability diagrams were measured as a function of barrier gates $V_3$ and $V_7$, while gate voltages $V_1$, $V_2$ and $V_8$ were tuned by the algorithm.
}
  \label{fig4}
\end{figure*}

We test the algorithm for different bias triangles measured on our device. Stability diagrams are measured as a function of barrier gate voltages $V_3$ and $V_7$, which are adjusted during centering of the bias triangles. The gate voltages optimised by the algorithm are $V_1$, $V_2$ and $V_8$. For simplicity, we chose to keep gate voltages $V_4$, $V_5$ and $V_6$ fixed. We checked their effect on the optimised bias triangles was weak. All measured stability diagrams are min-max normalised with respect to the stability diagram obtained after the initial tuning to the double quantum dot regime. In Fig.~\ref{fig4}, the optimisation of four different pair of bias triangles is shown. 

In cases 1 to 3, the initial bias triangles lack a well-defined shape, indicative of small inter-dot tunnel coupling~\cite{Vanderwiel2002}. Furthermore, pronounced co-tunnelling lines, which are denotative of second-order transport processes, are observed. As the optimisation progresses, the bias triangles separate from each other and acquire a sharper triangular shape. Also, co-tunnelling currents are reduced. In the fourth case, the initial stability diagram shows very faint bias triangles. The optimisation algorithm proves capable of increasing the current flowing through the double quantum dot while preserving most of the other bias triangle characteristics. More examples of bias triangle optimisations achieved by our algorithm can be found in the Supplementary Material.

\begin{figure}[ht]
\centering
  \includegraphics[width=\linewidth]{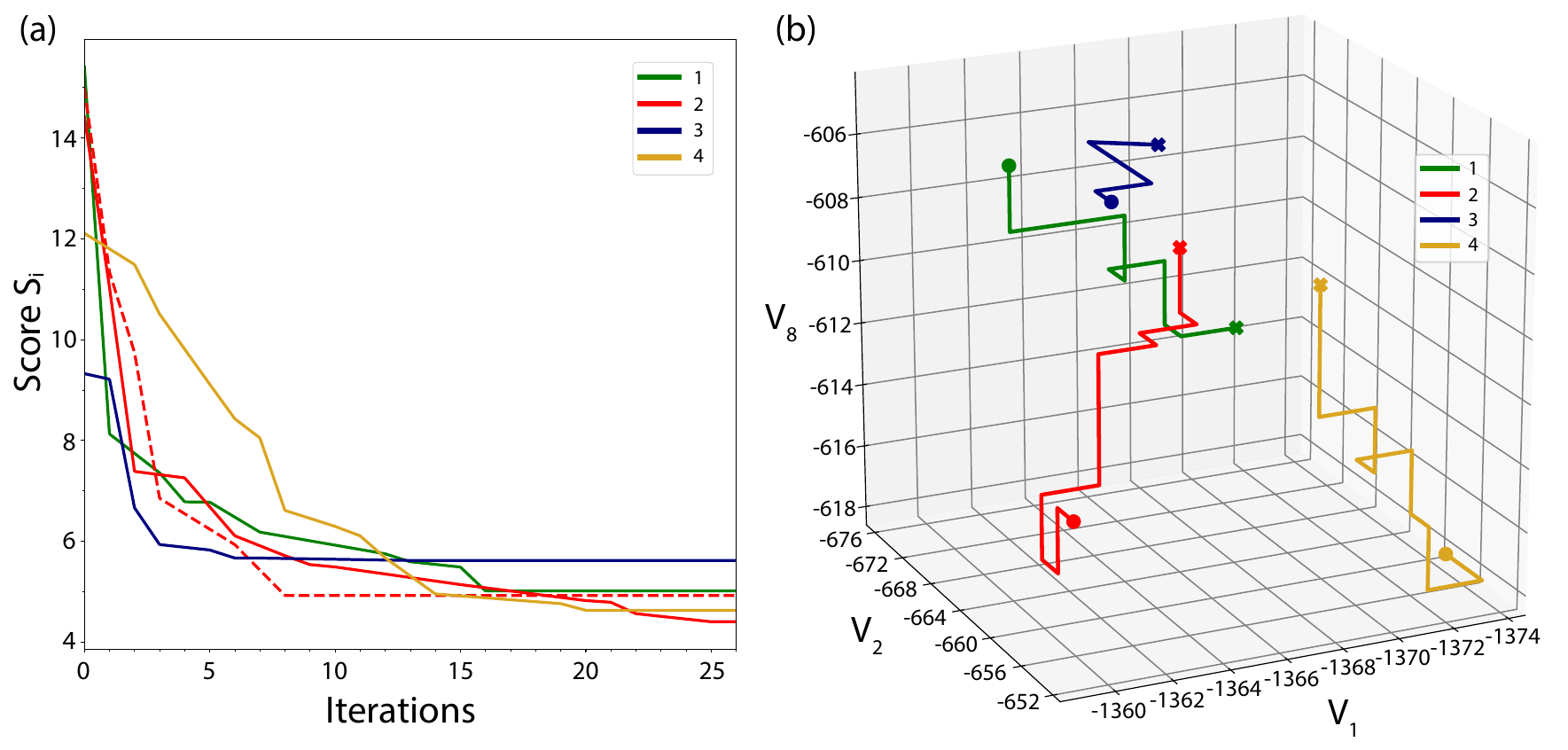}
  \caption{Score $S_i$ and gate voltage space trajectories during fine-tuning. (a) Score $S_i$ as a function of the number of iterations of the optimisation algorithm. Solid lines 1 to 4 correspond to the optimisation cases presented in Fig.~\ref{fig4}. The dashed line represents a different run of the algorithm for case 2. (b) Gate voltage space trajectories for the optimisation cases in Fig.~\ref{fig4}. The starting (final) gate voltage configuration is denoted by a circle (cross).}
  \label{fig5}
\end{figure}

Fig.~\ref{fig5}a shows $S_i$ as a function of the number of iterations of our algorithm for cases 1 to 4. Most of the optimisation takes place during the first ten iterations, after which the score does not change significantly. In all cases, the algorithm completes the optimisation within 26 iterations, corresponding to a total tuning time of 36 min. This time is limited by the measurement time, which could be drastically reduced by radio-frequency reflectometry techniques~\cite{Petersson2010,Ares2016,Schupp2018,Barthel2009,Pakkiam2018,West2019,Urdampilleta2019,Zheng2019}. 

In Fig.~\ref{fig5}b we plot the trajectories in gate voltage space corresponding to each optimisation case. The average distance in gate voltage space between the initial gate voltage configurations is greater than for the final gate voltage configurations. This suggests that there exists a region in gate voltage space for which the bias triangles exhibit the most favourable transport characteristics, regardless of their values of $V_3$ and $V_7$. Additional data can be found in the Supplementary Material.

\section{Conclusion}

We experimentally demonstrate an optimisation algorithm for the fine-tuning of bias triangles in gate-defined quantum dots. The algorithm scores real-time measurements by computing distances in the embedding space of a VAE. We show that this score can be used to locally optimise double quantum dot parameters in a completely automated manner. No prior knowledge of the device is required and the algorithm proves capable of tuning multiple device parameters at once. 

The robustness and efficiency of the decision model could potentially be improved by using Bayesian optimisation or reinforcement learning for proposing new voltage configurations and minimising the score. Also, while we utilised the Euclidean distance between two Gaussian distributions for computing scores, recent work argues that the decoder induces a Riemannian metric in the latent space~\cite{Arvanitidis2018}. This would imply that shortest paths in latent space do not correspond to straight lines. Therefore, it might prove insightful to implement a Riemannian metric to measure latent space distances. Finally, the influence of selecting targets with different characteristics, such as different excited state energies, could be investigated in the future.

While all measurements presented are performed on a gate-defined GaAs double quantum dot, the VAE was trained on data obtained from a Ge/Si core-shell nanowire device, showing the algorithm is readily applicable to different types of devices. Moreover, our algorithm can be adapted to include any number of additional gate electrodes, paving the way for the tuning of quantum dot arrays.  

\begin{acknowledgments}
We acknowledge discussions with E. A. Laird. This work was supported by the Royal Society, the EPSRC National Quantum Technology Hub in Networked Quantum Information Technology (EP/M013243/1), the Quantum Technology Capital grant (EP/N014995/1), Nokia, Lockheed Martin, the Swiss NSF Project 179024, the Swiss Nanoscience Institute and the EU H2020 European Microkelvin Platform EMP grant No. 824109. This publication was also made possible through support from Templeton World Charity Foundation and John Templeton Foundation. The opinions expressed in this publication are those of the authors and do not necessarily reflect the views of the Templeton Foundations. We acknowledge J. Zimmerman and A. C. Gossard for the growth of the AlGaAs/GaAs heterostructure. Lastly, we acknowledge F.N.M. Froning and F.R. Braakman for providing the Ge/Si training data. 

\end{acknowledgments}

%

\onecolumngrid

\beginsupplement
\clearpage
\section*{Supplementary Material}

\subsection{Latent space}

\begin{figure}[ht]
\centering
  \includegraphics[width=1\linewidth]{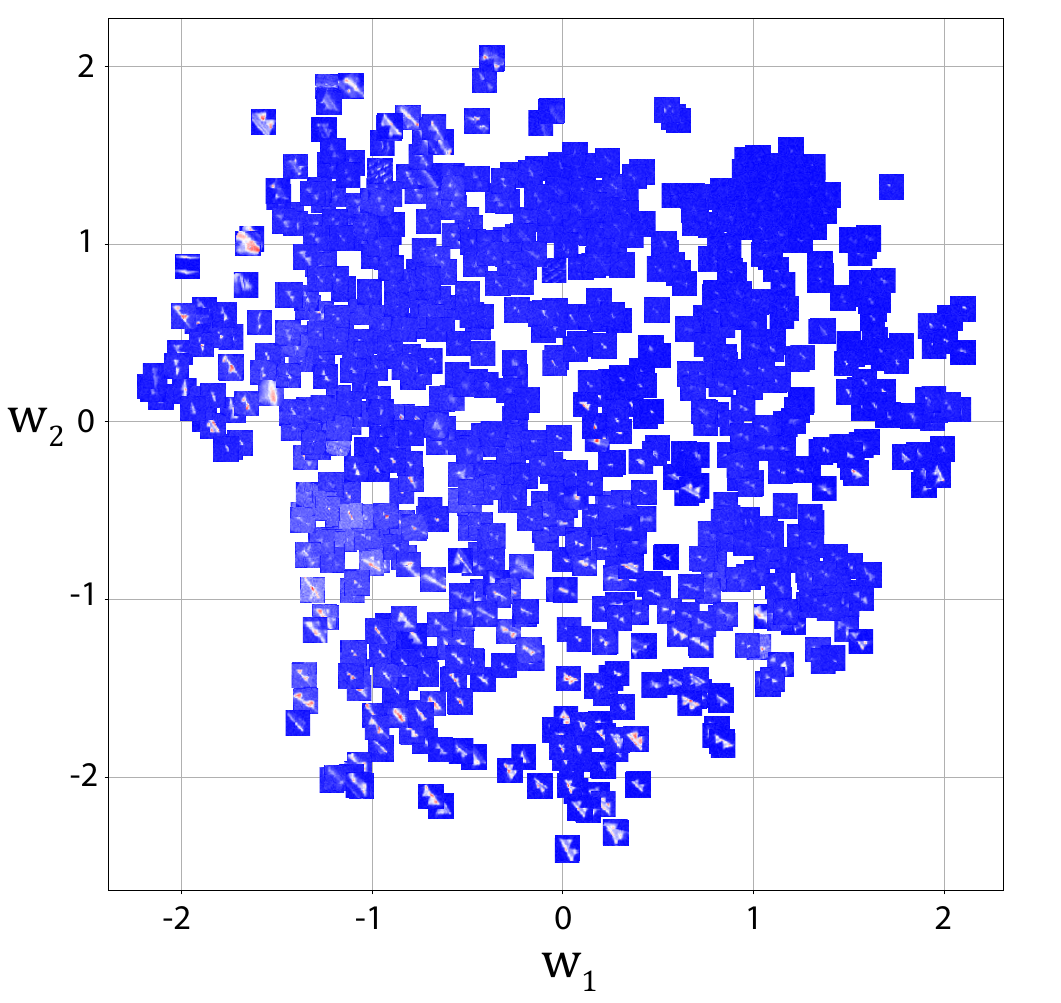}
  \caption{Latent space of the trained VAE. In order to visualise the ten-dimensional latent embeddings, t-SNE~\cite{vanDerMaaten2008} is applied for dimensionality reduction. The new two-dimensional latent space is described by a vector $\bm{w}$. The original training inputs are plotted at the embedding locations.}
  \label{fig:supp_latent}
\end{figure}

A representation of the latent space of the trained VAE with the bias triangles corresponding to each embedding is shown in Fig.~\ref{fig:supp_latent}.

\subsection{Optimisation}

\begin{figure}[ht]
\centering
  \includegraphics[width=0.8\linewidth]{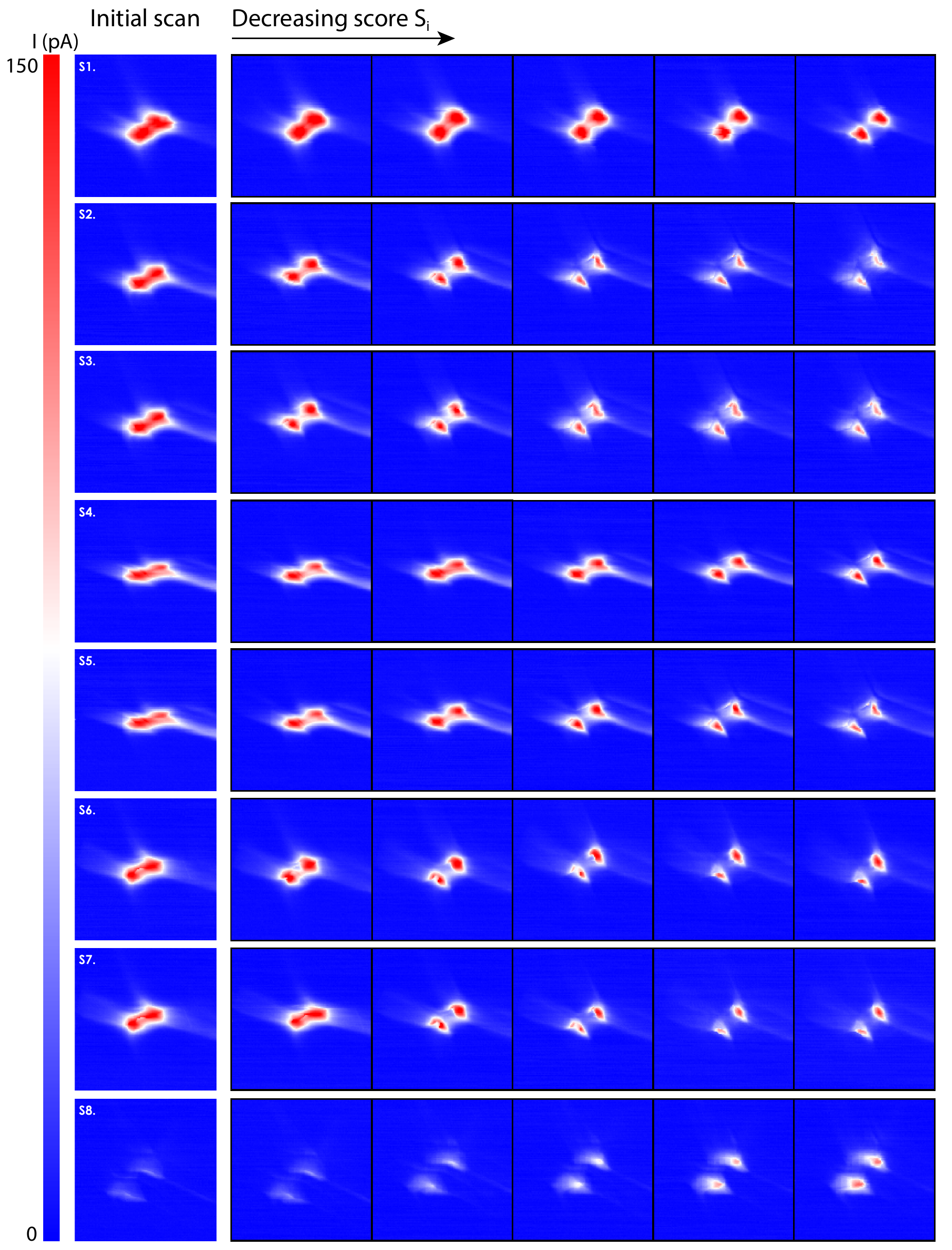}
  \caption{Stability diagrams of bias triangles at different iterations of the optimisation algorithm. The stability diagrams correspond to iterations for which the gate voltage configuration led to a decrease in score $S_i$. Only a selection of the bias triangle measurements at accepted gate voltage configurations are plotted.}
  \label{fig:supp_runs_1}
\end{figure}

\begin{figure}[ht]
\centering
  \includegraphics[width=0.8\linewidth]{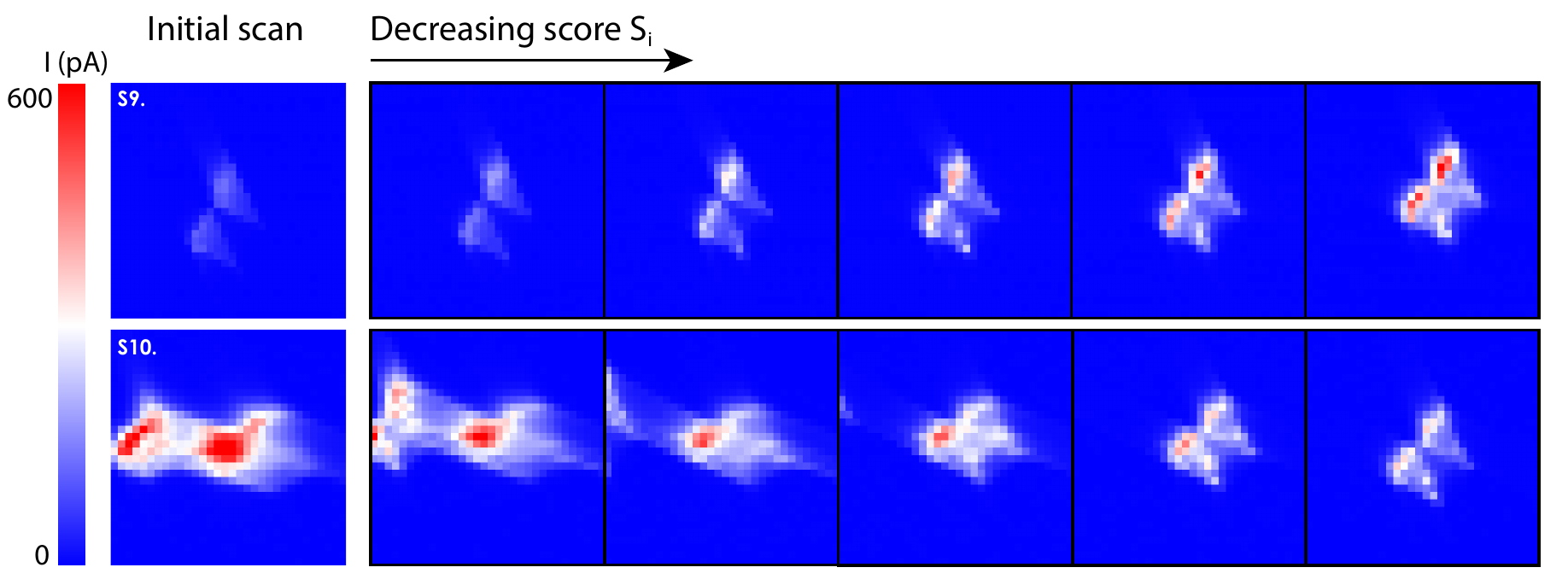}
  \caption{Stability diagrams of bias triangles at different iterations of the optimisation algorithm. The stability diagrams correspond to iterations for which the gate voltage configuration led to a decrease in score $S_i$. Only a selection of the bias triangle measurements at accepted gate voltage configurations are plotted.}
  \label{fig:supp_runs_2}
\end{figure}

\begin{figure}[ht]
\centering
  \includegraphics[width=0.6\linewidth]{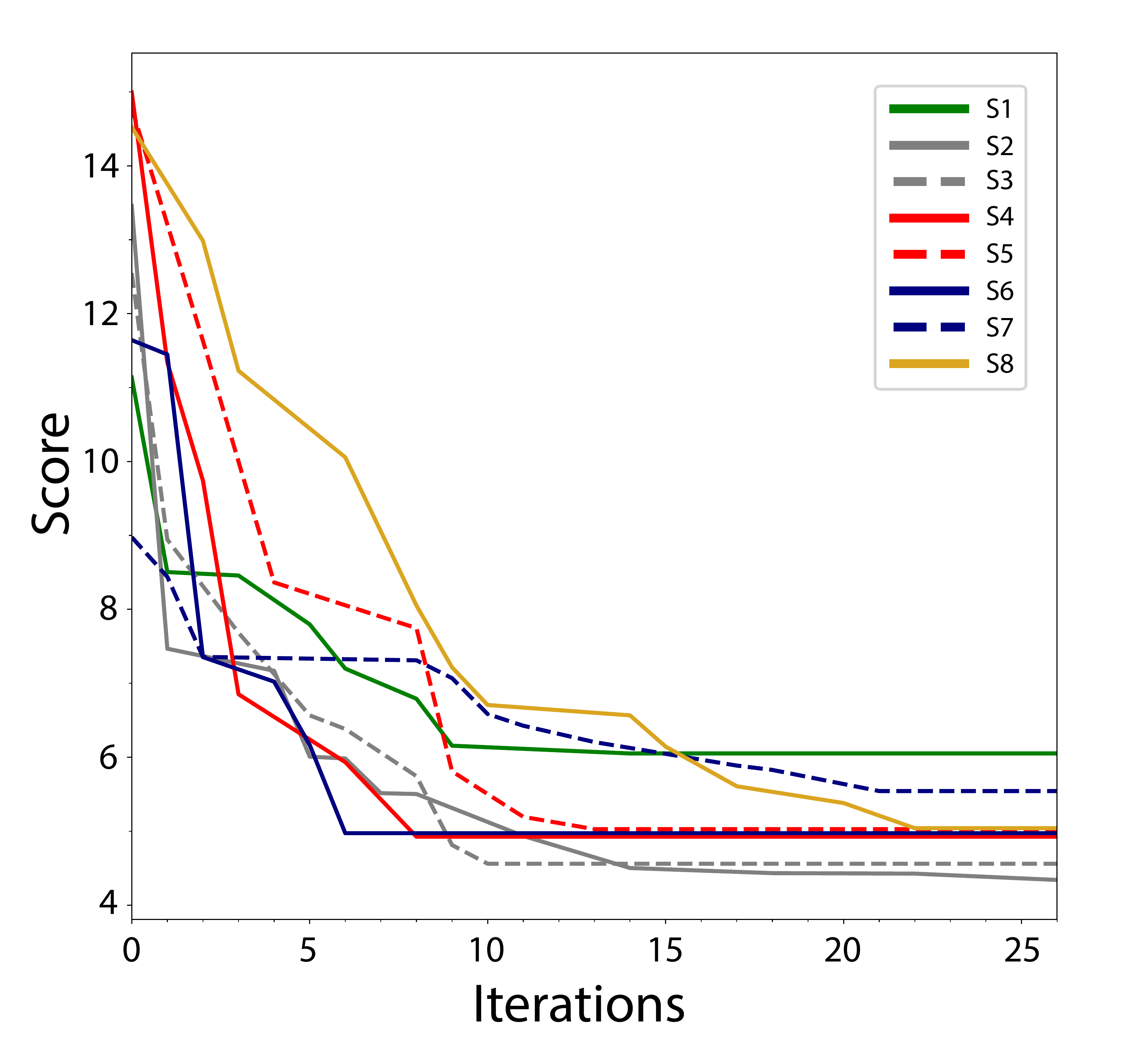}
  \caption{Score $S_i$ as a function of the number of iterations of the optimisation algorithm. The indexed lines correspond to the optimisation cases presented in Fig.~\ref{fig:supp_runs_1}. Dashed lines of the same colour represent different runs of the optimisation algorithm for given cases.}
  \label{fig:supp_score_it}
\end{figure}

The optimisation of different pairs of bias triangles is shown in Fig.~\ref{fig:supp_runs_1} (cases S1 to S8) and Fig.~\ref{fig:supp_runs_2} (cases S9 and S10). The stability diagrams correspond to gate voltage configurations for which a decrease in score $S_i$ is observed. Fig.~\ref{fig:supp_score_it} presents the VAE score $S_i$ as a function of the number of iterations of the optimisation algorithm for the optimisation cases in Fig.~\ref{fig:supp_runs_1} and Fig.~\ref{fig:supp_runs_2}. The total gate voltage changes during fine-tuning are presented in Table~\ref{table1}. 

\begin{table}
\caption{\label{tab:example}Total gate voltage change (mV) during fine-tuning. Gates voltages $V_1$, $V_2$ and $V_8$ were optimised, whereas gates $V_3$ and $V_7$ were used to center the bias triangles.}
\begin{tabular}{llllll}
\hline
Case &  $\Delta V_1$    &  $ \Delta V_2$  &  $ \Delta V_3$  &  $ \Delta V_7$ &  $\Delta V_{8}$ \\
\hline
1   &  -8.0   & 0  & 6.29  & 5.92 & -6.0  \\ 
 2   &  -6.0   & -8.0  & 6.57    & 0 & 6.0 \\
 3   &  -4.0   & -8.0  & 5.26    & 1.97 & 0 \\
 4   &  6.0   & 4.0  & -3.94    & -7.89 & 10.0 \\  
 S1   &  -10.0  & 0  & 5.26  & 4.60 & 0 \\         
S2   &  -4.0   & -10.0 & 5.92   & 1.31  & 0 \\
 S3   &  0   & -8.0 & 2.63 & 0 & -2.0 \\
S4   &  -6.0   & -2.0  & 3.94  & 1.97 & 2.0 \\
   S5   &  -4.0   & -6.0  & 4.60    & 0.66 & 2.0 \\
 S6   &  0   & -8.0  & 3.29    & -1.31 & 2.0 \\
   S7   &  -2.0   & -8.0  & 3.94    & 0.66 & 0 \\
 S8   &  14.0   & 0  & -7.23    & -9.86 & 4.0 \\
 \hline
 \label{table1}
\end{tabular}
\end{table}



\end{document}